# New Concept for Internal Heat Production in Hot Jupiter Exo-Planets

**J. Marvin Herndon**
**Transdyne Corporation**
**San Diego, CA 92131 USA**

December 20, 2006

Communications: mherndon@san.rr.com  http://UnderstandEarth.com

**Abstract**

Discovery of hot Jupiter exo-planets, those with anomalously inflated size and low density relative to Jupiter, has evoked much discussion as to possible sources of internal heat production. But to date, no explanations have come forth that are generally applicable. The explanations advanced typically involve presumed tidal dissipation and/or converted incident stellar radiation. The present, brief communication suggests a novel interfacial nuclear fission-fusion source of internal heat production for hot Jupiters that has been overlooked by theoreticians and which has potentially general applicability.

**Introduction**

Astronomical observations of planets orbiting stars other than our Sun will inevitably lead to a more precise understanding of our own Solar System and as well, perhaps, of the Universe as a whole. For that to occur, one must be willing to consider exo-planet observations objectively, which may mean re-examining astrophysical concepts previously thought to be on secure footing. But, of course, doing that is just good science.

The discovery of so-called "hot Jupiter" exo-planets, those with anomalously inflated size and low density relative to Jupiter, has evoked much discussion as to possible sources of internal heat production. But to date, no explanations have come forth that are generally applicable. For example, hot Jupiters are found with insufficient eccentricity to be heated internally by tidal dissipation as originally suggested by Bodenheimer, Lin, & Mardling (2001). Other ideas, such as internal conversion of incident radiation into mechanical energy (Showman & Guillot 2002) and on-going tidal dissipation due to a non-zero planetary obliquity (Winn & Holman 2005) also appear to lack general applicability. Charbonneau et al. (2006), note that two cases [HD 209458b and HAT-P-1b] suggest at least "…there is a source of internal heat that was overlooked by theoreticians".

The purpose of this brief communication is to suggest a source of internal heat production for hot Jupiters that indeed has been overlooked by theoreticians and which has potentially general applicability.

# Planetocentric Nuclear Fission Reactors

In the late 1960s, astronomers discovered that Jupiter radiates into space about twice as much energy as it receives from the Sun. Later, Saturn and Neptune were also found to radiate prodigious quantities of internally generated energy. That excess energy production has been described by Hubbard (1990) as being "one of the most interesting revelations of modern planetary science." Stevenson (1978), discussing Jupiter, stated, "The implied energy source ... is apparently gravitational in origin, since all other proposed sources (for example, radio-activity, accretion, thermonuclear fusion) fall short by at least two orders of magnitude...." Similarly, more than a decade later, Hubbard (1990) asserted, "Therefore, by elimination, only one process could be responsible for the luminosities of Jupiter, Saturn, and Neptune. Energy is liberated when mass in a gravitationally bound object sinks closer to the center of attraction ... potential energy becomes kinetic energy ....."

In about 1990, when I first considered Jupiter's internal energy production, that explanation did not seem appropriate or relevant because about 98% of the mass of Jupiter is a mixture of hydrogen and helium, both of which are extremely good heat transport media. Having knowledge of the fossil natural nuclear fission reactors that were discovered in 1972 at Oklo, Republic of Gabon, in Western Africa, I realized a different possibility and proposed the idea of planetary-scale nuclear fission reactors as energy sources for the giant planets (Herndon 1992). At first I demonstrated the feasibility for thermal neutron reactors in part using Fermi's nuclear reactor theory, *i.e.*, the same calculations employed in the design of commercial nuclear reactors and used by Kuroda (1956) to predict conditions for the natural reactors that were later discovered at Oklo. Subsequently, I extended the concept to include planetocentric fast neutron breeder reactors, which are applicable as well to non-hydrogenous planets, especially the nuclear georeactor as the energy source for Earth's magnetic field (Herndon 1993, 1994).

There is strong terrestrial evidence for the planetocentric nuclear reactor concept. In the 1960's geoscientists discovered occluded helium in oceanic basalts which, remarkably, possessed a higher $^3$He/$^4$He ratio than air. At the time there was no known deep-Earth mechanism that could account for the $^3$He, so it was assumed that the $^3$He was a primordial component, trapped at the time of Earth's formation, which was subsequently diluted with $^4$He from radioactive decay. State-of-the-art numerical simulations of georeactor operation, conducted at Oak Ridge National Laboratory, yielded fission-product helium, as shown in Fig. 1, with isotopic compositions within the exact range of compositions typically observed in oceanic basalts (Herndon 2003; Hollenbach & Herndon 2001). For additional information, see Rao (2002).

One might expect planetocentric nuclear fission reactors to occur within exo-planets that have a heavy element component, provided the initial actinide isotopic compositions are appropriate for criticality. And indeed, planetocentric nuclear fission reactors may be a crucial component of hot Jupiter exo-planets. But it is unlikely that fission-generated heat alone would be sufficient to create the "puffiness" that is apparently observed. For example, as calculated using Oak Ridge nuclear reactor numerical simulation software, a one Jupiter-mass exo-planet without any additional core enrichment of actinide elements could produce a constant fission-power output of ~ 4 x $10^{21}$ ergs/s for only ~ 5 x $10^{8}$ years. Even with that unrealistically brief interval, the fission-power output is orders of magnitude lower than the $10^{26}$ to $10^{29}$ ergs/s hot Jupiter model-estimates made by Bodenheimer, et al. (2001).

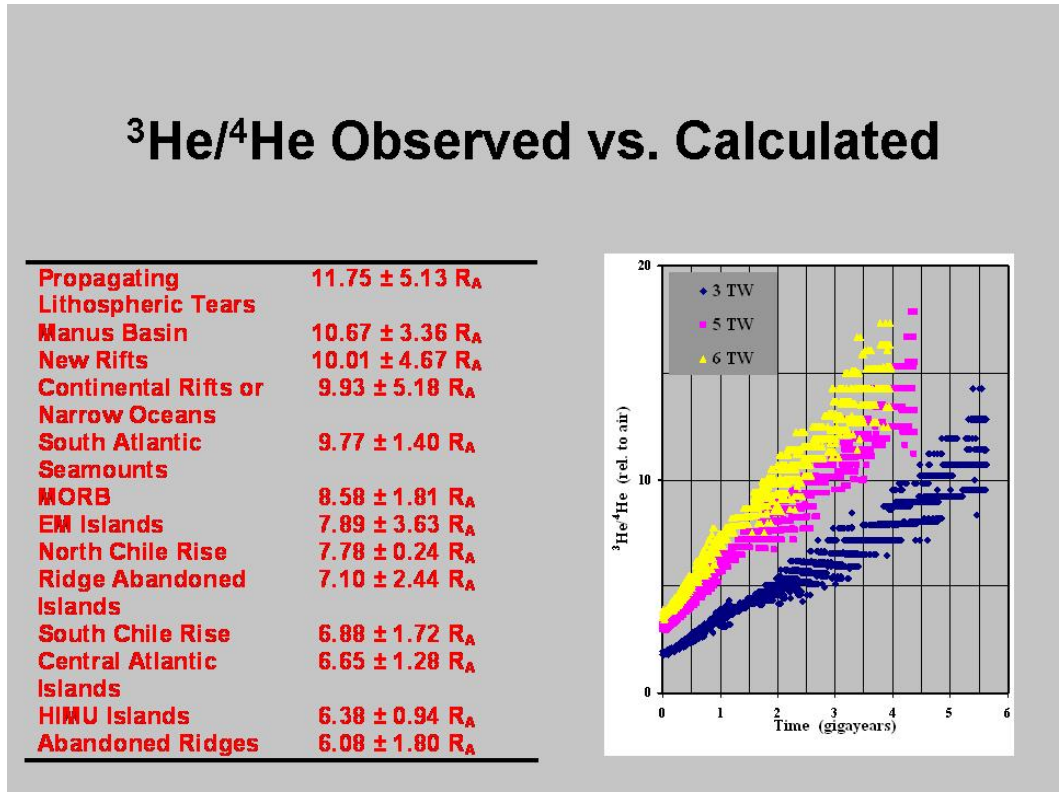

**Fig. 1.** Ratio of $^3$He/$^4$He, relative to that of air, tabulated for oceanic basalts at 95% confidence level shown for comparison with similar values obtained from nuclear georeactor numerical calculations. In particular, note the distribution of calculated values at 4.5 gigayears, the approximate age of the Earth. Adapted from Herndon (2003).

**Thermonuclear Fusion Ignition by Nuclear Fission**

At the beginning of the 20th century, understanding the nature of the energy source that powers the Sun and other stars was one of the most important problems in physical science. Initially, gravitational potential energy release during protostellar contraction was considered, but calculations showed that the energy released would only be sufficient to power a star for a few million years and life has existed on Earth for a longer time. The discovery of radioactivity and the developments that followed led to the idea that thermonuclear fusion reactions power the Sun and other stars.

Thermonuclear fusion reactions are called "thermonuclear" because temperatures on the order of a million degrees Celsius are required. The principal energy released from the detonation of hydrogen bombs comes from thermonuclear fusion reactions. The high temperatures necessary to ignite H-bomb thermonuclear fusion reactions comes from their A-bomb nuclear fission triggers. Each hydrogen bomb is ignited by its own small nuclear fission A-bomb.

By 1938, the idea of thermonuclear fusion reactions as the energy source for stars had been reasonably well developed (Bethe 1939), but nuclear fission had not yet been discovered (Hahn & Strassmann 1939). Astrophysicists assumed that the million-degree-temperatures necessary for stellar thermonuclear ignition would be produced by the in-fall of dust and gas during star formation and have continued to make that assumption to the present, although clearly there have been signs of potential trouble with the concept. Proto-star heating by the in-fall of dust and gas is off-set by radiation from the surface which is a function of the fourth power of temperature. Generally, in numerical models of protostellar collapse, thermonuclear ignition temperatures, on the order of a million degrees Celsius, are not attained by the gravitational in-fall of matter without assumption of an additional shockwave induced sudden flare-up (Hayashi & Nakano 1965; Larson 1984) or result-optimizing the model-parameters, such as opacity and rate of in-fall (Stahler et al. 1994).

After demonstrating the feasibility for planetocentric nuclear fission reactors, I suggested that thermonuclear fusion reactions in stars, as in hydrogen bombs, are ignited by self-sustaining, neutron induced, nuclear fission (Herndon 1994). The idea that stars are ignited by nuclear fission triggers opens the possibility of stellar non-ignition, a concept which may have fundamental implications bearing on the nature of dark matter (Herndon 1994) and dark galaxies (Herndon 2006). I now suggest the possibility that hot Jupiter exo-planets may derive much of their internal heat production from thermonuclear fusion reactions ignited by nuclear fission.

Unlike stars, hot Jupiter exo-planets are insufficiently massive to confine thermonuclear fusion reactions throughout a major portion of their gas envelopes. One might anticipate instead fusion reactions occurring at the interface of a central, internal substructure, presumably the exo-planetary core, which initially at least was heated to thermonuclear ignition temperatures predominantly by self-sustaining nuclear fission chain reactions. After the onset of fusion at that reactive interface, maintaining requisite thermonuclear-interface temperatures might be augmented to some extent by fusion-produced heat, which would as well expand the exo-planetary gas shell, thus decreasing the exo-planet's density.

Through good science, resourcefulness, and hard work, exo-planetary astronomers have opened a new window of opportunity to understand the nature of remote planetary systems as well as our own Solar System. We, as scientists attempting to understand those observations and the inferences that might result there from, have the responsibility to be objective, perspicacious, imaginative, and open-minded. This brief communication is presented in that spirit.